\documentstyle[twocolumn,aps]{revtex}
\begin{document}


\title{Monte Carlo Simulations for the Magnetic Phase Diagram of
the Double Exchange Hamiltonian}

\author{M.J. Calder\'on and L. Brey}
\address{Instituto de Ciencia de Materiales (CSIC).
Campus de la Universidad Aut\'onoma,
28049, Madrid, Spain.  }

\date{\today}

\maketitle

\begin{abstract}
We have used Monte Carlo simulation techniques to obtain
the magnetic phase diagram of the double exchange Hamiltonian.
We have found that the Berry phase of the hopping amplitude
has a negligible effect in the value of the magnetic critical 
temperature. 
To avoid finite size problems
in our simulations
we have also developed
an approximated expression for the double exchange energy. This
allows us to obtain the critical temperature for the 
ferromagnetic to paramagnetic transition more accurately. 
In our calculations we do not observe any strange behavior
in the kinetic energy, chemical potential or electron density of
states near the magnetic critical temperature. Therefore, we conclude
that other effects, not included in the double exchange Hamiltonian,
are needed in order to understand the metal-insulator transition
which occurs in the manganites.
\end{abstract}
PACS number 71.10.-w, 75.10.-b
\section{Introduction}
 
  The recent discovery of colossal magnetoresistance (CMR)\cite{prim}
in mixed-valence compounds of the form 
$La ^{3+} _{1-x} A ^ {2+} _{x} Mn ^{3+} _{1-x} Mn ^{4+} _{x} O _{3} ^{2-}$,
(where $A$ can be Ca, Sr or Ba) has revived the interest in these
materials with perovskite structure.
The interest is focused on the phase diagram and the magneto-transport 
properties. Both features have been determined experimentally\cite{rvw}. 
For $0.1 \leq x \leq 0.5$ and low temperatures the system is metallic and
presents ferromagnetic order.
As the temperature increases the system becomes insulator and paramagnetic.
The ferro-paramagnetic transition occurs at an
$x$ dependent critical temperature $T_c (x) \sim 300K$. The 
colossal magnetoresistance effects are observed for temperatures close to
$T_c (x)$. 
For $x \leq 0.1$ and low temperatures the system is a 'layer'
antiferromagnetic with ferromagnetic coupling inside planes.

In these compounds
the electronically active orbitals are the $Mn$ $d$ orbitals.
The number of $d$ electrons per $Mn$ is $x+3$, namely, four electrons
per $Mn^{3+}$ and three per $Mn^{4+}$. The cubic symmetry
and the strong Hund's rule coupling make that three electrons get
trapped in the $t_{2g}$ states and therefore these electrons become
electrically inert, forming a core spin $S$ of magnitude $3/2$.
The rest of the electrons go to the $e_g$ orbitals and  
they  get strongly coupled to $S$ by the  Hund's rule
coupling. Consequently, all the spins on each $Mn$ prefer to be parallel.
For small values of $x$, the perovskites show a long range Jahn-Teller
order which selects a preferred $d$ orbital and therefore it is possible
to assume that the electrons move only through one $d$ orbital.

The double exchange (DE) mechanism developed by Zener
\cite{zener,anderson,degennes},
explains the existence of ferromagnetism and metallic behavior at low
temperatures. In this model, the electrons get mobility between the manganese
ions using oxygen, which is magnetically inert, as an intermediate.
As a consequence, the tunneling takes place between two
configurations in which the $Mn$ ions of different charge
($Mn ^{3+}$ and $Mn ^{4+}$) interchange their valence states.
This conduction process is proportional to the electron transfer integral
$t$ if the core spins of $Mn ^{3+}$ and $Mn ^{4+}$ are aligned.
Otherwise, the transfer integral is inversely proportional to the Hund's
rule coupling energy, which for the $Mn$ ions is much larger than $t$. 
In the DE model the ferromagnetism  is then induced via this electron 
conduction process.

 In the limit of large Hund's coupling the spin of the electron in the
active $d$ orbital is parallel to the local core spin ${\bf S} _i$, which 
is treated as a classical rotator with the normalization $\mid {\bf S} \mid =1$
and characterized by the angles $\theta _i$ and
$\phi _i$. Then, semiclassically, the effective hopping Hamiltonian is 

\begin{equation}
H= - \sum _{\langle i,j\rangle} \left ( t _{i,j} C^ {+} _i C _ j + h.c. \right )
\, \, \, \, \, .
\label{hcomplex}
\end{equation}
Here $C^ {+} _i$ creates an electron at site $i$ with spin parallel 
to ${\bf S} _i$, $\langle i,j\rangle$ denotes the nearest-neighbor pairs,
which are  only counted once, 
and the hopping amplitude acquires a Berry's phase 
and it becomes a complex number given 
by\cite{muller},
\begin{equation}
t_{i,j}= t \left ( \cos {{\theta _ i } \over 2}
\cos {{\theta _ j } \over 2}
-\sin {{\theta _ i } \over 2}
\sin {{\theta _ j } \over 2}
e ^ { i ( \phi _i - \phi _j )} \right ) \, \, \, .
\label{tcomplex}
\end{equation}
This complex hopping appears after rotating the conduction electron spins
so that the spin quantization axis at site $i$
is parallel to  ${\bf S} _i$, and then project onto the spin parallel
to ${\bf S} _j$. 
In this paper we will call complex DE (cDE)  model the system governed by the 
Hamiltonian (\ref{hcomplex}) with the hopping given by (\ref{tcomplex}).

For topologies where electrons close paths do not occur, it is
possible to choose wave functions phase factors such that

\begin{equation}
t_{i,j} \rightarrow \mid t_{i,j}\mid =t \cos { \theta _{i,j} \over 2}
\, \, \, \, ,
\label{treal}
\end{equation}
being
$\theta _{i,j}$ the angle between the semiclassical spins ${\bf S} _i$
and ${\bf S} _j$.

%

In this work we will use the name  DE model for the system controlled by
equation (\ref{hcomplex}) with the hopping (\ref{treal}). 
This is the Hamiltonian proposed by Anderson and Hasegawa\cite{anderson}
 as a generalization
of the Hamiltonian describing the tunneling between two fixed sites
$i$ and $j$.

De Gennes\cite{degennes} and Kubo and Ohata\cite{kubo} calculated,
in a mean field theory approximation, the magnetic phase diagram of the DE model.
Their results show a ferro-paramagnetic phase transition at a
critical temperature
\begin{equation}
{{T^{MF} _c } \over t}
= 1.6 \langle C^ {+} _i C _ j \rangle _0 \, \, \, 
\label{tmf}
\end{equation}
This transition is accompanied by a change in the temperature dependence
of the resistivity.
In equation (\ref{tmf}), $\langle \hat{O} \rangle _0$ means the expectation value of the operator
$\hat{O}$ at zero temperature.

However, recently it has been pointed out\cite{millis1}
that the critical temperatures predicted by the mean field theory 
of the DE model
is much bigger than the observed experimentally.
Furthermore, the resistivity implied
by the hopping (\ref{treal}) is incompatible with many aspects of the experimental 
information.
Millis {\it et al}\cite{millis1} proposed that in 
addition to the DE mechanism a strong electron-phonon interaction plays a 
crucial role in these perovskites. The interplay between these two 
effects could reproduce the experimental critical temperature and the observed
behavior of the resistivity at the magnetic transition 
temperature\cite{roder,millis2}.

On the other hand, Yunoki {\it et al}\cite{yunoki}  claim that the inclusion of the
phase in the hopping term can lower the critical temperature
down to the experimental values.
Also, for large but finite Hund's coupling, they obtain that at low 
concentrations phase separation between hole-poor antiferromagnetic
and hole-rich ferromagnetic regions occurs.
The existence of phase separation is 
also  obtained by solving the DE model plus
an  antiferromagnetic coupling between next neighbors
core spins\cite{paco1}.

In this paper we are interested in two points; first, in comparing
the magnetic phase diagram of the two DE models, equations  
(\ref{tcomplex}) and (\ref{treal}),
and second, in obtaining an approximated expression for the DE  energy
which allows us to perform Monte Carlo calculations in bigger unit cells
and to obtain accurate values of the magnetic critical temperature.
The main results of our calculations are  shown in Fig.\ref{tc}.  Here we show
the critical temperature versus concentration, for the complex DE model and
for the real DE model, in the case of a unit cell of size $4 \times 4 \times 4$.
We obtain that there is not big difference between the critical temperatures
of both models.
This result indicates that the contribution to the partition function 
of electronic configurations in 
which fermions move on closed loops in real space can be neglected.
In Fig.\ref{tc} we also show Monte Carlo results obtained by using
an approximated expression for the electron kinetic energy.
We have checked in small systems that this approximation
is realistic. From our results we obtain that the critical temperature of the
DE model is in the range of the experimental one\cite{rvw}.

The paper is organized as follows: in section II we describe briefly the 
Monte Carlo algorithm we use and in section III we present the results
obtained for the cDE and DE Hamiltonians. In section IV we introduce an approximation
to the DE energy which allows us to perform Monte Carlo
simulations in  bigger systems. Section V is devoted to show the results obtained
in the frame of this approximation. Finally, we present the conclusions 
in section VI.

\section{Description of the Monte Carlo algorithm}

For calculating the magnetic phase diagrams, we have performed classical 
Monte Carlo (MC) simulations on the classical core spin angles.
The simulations are done in $N \times N \times N$ cubic lattices
with periodic boundary conditions.
Although the localized spins are considered classical, the kinetic energy of the 
conduction electrons is calculated by diagonalizing the DE Hamiltonian because
we consider it as a quantum quantity.

The standard Metropolis algorithm\cite{binder} was used in the MC
simulations.
The sites to be considered for a change in the spin orientation 
are randomly chosen. Once a site is selected for a spin reorientation,
the angle associated with an attempted change of the spin is 
chosen at random from within a specified range\cite{serena}.
Then the energy change, $\Delta E$, associated with the attempted
update, is calculated. If
the quantity $exp(- \Delta E / T)$ is smaller than a random number between
$0$ and $1$, the change is allowed, otherwise, it is rejected.
Typically, 5000-7000 MC steps per spin are used for equilibration
and 3000-5000 steps for spin are used for calculating averages.

In the simulations we calculate the average of the internal energy, $E$,
and the average of the absolute value of the magnetization, $M$,
\begin{equation}
E = {1 \over { N ^3}} \langle H \rangle 
\, \, \, \, ,
\end{equation}
\begin{equation}
M =  {1 \over { N ^3 } } \langle  \left | \sum _i {\bf S _i} \right | \rangle
\, \, \, \, \, \, ,
\end{equation}
where $\langle \, \, \rangle$ denotes statistical average. 
Since the MC updating procedure generates uniform
rotations of the spin system, a calculation of the MC
average of the direction of the magnetization  is not meaningful.
Also we calculate the average value of the width of the 
density of states, $W$, and the value of the chemical potential, $\mu$, 
measured with respect to the bottom of the density of states,
\begin{eqnarray}
W & = & \langle \varepsilon _{max} - \varepsilon _{min} \rangle
\, \, \, \, \\
\mu & = & \langle \varepsilon _{occ} - \varepsilon _{min} \rangle
\, \, \, \, ,
\end{eqnarray}
where $ \varepsilon _{min}$ and $\varepsilon _{max}$ are 
the minimum and the maximum energies  obtained from diagonalizing the
electron Hamiltonian, and $\varepsilon _{occ}$ is the higher energy of the 
occupied states.

\section{Monte Carlo results for the double exchange models}

In order to obtain the internal energy in 
each of the Monte Carlo steps, it is necessary to diagonalize the electronic
Hamiltonian. The diagonalization is very expensive in terms
of CPU time, and therefore we can only study small size systems.
Here we present  the results obtained for a cubic unit cell of size 
$N=4$ and for the electron concentrations $x=0.1$, $0.25$ and
$0.3$.

In figure \ref{de-cde}, it is presented, for the DE and the cDE Hamiltonians, 
the magnetization, $M$, as a function of the temperature, $T$. 
Due to the finite size of the unit cells used in the simulations,
$M$ is different from zero at any temperature, and we define
the critical temperature, $T_c$, as the point where the second derivative
of $M$ with respect $T$ changes sign.
This way of obtaining the critical temperatures, implies uncertainties 
of around $10\%$ of the value of $T_c$. 
In any case we obtain that the critical temperatures of the DE and the cDE
Hamiltonians are practically the same. 
In figure \ref{E-t} we plot the internal energy $E$ as a function of $T$.
Note that the difference between the internal energies  obtained by using
the cDE and using the DE Hamiltonians is rather small. That is the reason
why both models give very similar magnetic critical temperatures. 

These  results imply that 
in order to calculate total kinetic energies
the close paths
are almost irrelevant. 

In figure \ref{W-mu-t} we plot, as a function of the temperature, the width
of the density of states $W$ and the chemical potential $\mu$, for the
DE case.
Note that at temperatures close to $T_c$, the
bandwidth and the chemical potential have a value bigger than the obtained
in the  fully disordered case, $T \rightarrow \infty$. 
This is because in the DE models the ferromagnetic to paramagnetic 
transition is a second order phase transition, and therefore
the internal energy is continuous at the critical temperature.
Moreover, the bandwidth and the chemical potential present 
a continuous behavior near $T_c$.
In the DE models we do not observe any change in the
electronic states  at the  ferro-paramagnetic critical temperature.

\section{Approximation for the double exchange energy}

In the MC calculations of the DE models, the size of the matrix to diagonalize 
impose a restriction on the dimension of the unit cell used in the simulation.
In order to be able to perform  simulations in bigger systems, we have 
developed a second order perturbation theory for obtaining
an expression for the electron 
kinetic energy 
in a background of randomly oriented core spins.

We start writing the Hamiltonian (\ref{hcomplex}) in the form,
\begin{equation}
H   \equiv    H _0 + V 
\end{equation}
with

\begin{eqnarray}
 H_0  & \equiv & - \bar {t} \sum _{ \langle i,j\rangle} \left ( C ^+ _i C _j + h.c. \right ) 
\\
 V  & \equiv & - \sum _{ \langle i,j\rangle} \left ( \delta t _ {i,j} C ^+ _i C _j + h.c. \right )
\, \, \, ,
\end{eqnarray}
being $\bar {t}$ the average of the absolute value of the hopping amplitude,
\begin{equation}
\bar {t} = { 1 \over {3 N _0}} \sum _ { \langle i,j\rangle} \mid t _ {i,j}\mid \, \, \, ,
\end{equation}
and 
\begin{equation}
\delta t _ {i,j}  = t _ {i,j}- \bar {t} \, \, \,.
\end{equation}
Here $N_0$ is the total number of $Mn$ ions in the system.
Note that since the Hamiltonian is hermitic, $\bar {t}$ is a real
quantity.

Given a disordered system, characterized by a set of $\{ t_{i,j}  \}$, 
we want to obtain the expectation value of $H$.
The Hamiltonian $H_0$ can be diagonalized by using the Bloch states,
\begin{equation}
\mid {\bf k} \rangle = { 1 \over { \sqrt{N_0} } } \sum _ i 
e ^{ i {\bf k} \cdot {\bf R_i } }
\mid i\rangle \, \, \, \, ,
\end{equation}
where $\mid i\rangle$ represents the atomic orbital at site $i$, 
${\bf R_i }$ are the lattice vectors
and ${\bf k}$ is a wave vector in the first Brillouin zone.
The energy of the state $\mid{\bf k}\rangle$ is 
\begin{equation}
\varepsilon ({\bf k}) = -2 \bar {t} \left (
\cos { k_x a } +
\cos { k_y a } +
\cos { k_z a } \right ) \, \, \, ,
\end{equation}
being $a$ the lattice parameter.
In function of these eigenvalues, the expectation value of $H_0$ is,
\begin{equation}
E_0 = \sum _ {k} ^{occ} \varepsilon ({\bf k}) \, \, \, ,
\end{equation}
where the sum is over the occupied states. 
From the value of $E_0$ we can obtain the expectation value of
$C^+ _i C _j$,
\begin{equation}
\langle C^+ _i C _j\rangle_0 = - {{E_0} \over { 6 \bar {t} N _0 }} 
\, \, \, ,
\end{equation}
which is independent of the value of $ \bar {t}$.

To obtain the expectation value of $\langle V \rangle$, we first notice that since  the set 
$ \{ \delta t _{i,j} \}$ are randomly distributed,
$\langle {\bf k}\mid V\mid {\bf k }\rangle =0$, and it is necessary second order
perturbation theory 
in order to get a correction to the
expectation value of $H_0$, 
\begin{equation}
E _2 \simeq - \sum _ {\bf k} ^ {occ} \langle {\bf k} \mid V G_0 ( \varepsilon _{\bf k} ) V
\mid {\bf k } \rangle 
\, \, \, \, ,
\end{equation}
where 
\begin{equation}
G_0 ( \hbar \omega ) = \sum _{\bf k}  { 
{ \mid{\bf k} \rangle \langle {\bf k}\mid } \over { \hbar \omega - \varepsilon _{\bf k} }}
\end{equation}
is the Green function of the perfect crystal. Because the values 
$ \{ \delta t _{i,j} \}$  are  not correlated, it is easy to obtain the expression,
\begin{equation}
E _2 = - a_2 
 \sum _ { \langle i,j\rangle} 
(\bar {t} - t _ {i,j}) ^ 2 \, \, \, ,
\end{equation}
with 
\begin{equation}
a_2 =  -{1 \over {2 N _0  ^2}} \sum _{\bf k} ^ {occ} 
\sum _{ {\bf k}  ^{\prime}} 
\left ( 6 + { { \varepsilon ( {\bf k}  +  {\bf k}  ^{\prime} )}  \over {\bar {t}}}
\right ) 
{ 1 \over { \varepsilon ( {\bf k}) -\varepsilon ( {\bf k}^{\prime})}}
\, \, \, \, .
\end{equation}

Adding  
$E_0$ and $E_2$,  
we obtain the following expression
for the energy of the system,
\begin{equation}
 E  \simeq 
-2 \langle C ^ + _i C _j \rangle _0  \sum _{\langle i,j\rangle} \cos { { \theta _{ij}} \over 2} -
 a_2 
 \sum _ { \langle i,j\rangle} 
(\bar {t} - t _ {i,j}) ^ 2 \, \, \, .
\label{energy-approx}
\end{equation}

In figure \ref{coef} we plot  $\langle C ^ + _i C _j \rangle _0 $ and
$a_2$ as a function of the electron density $x$. 
The first term in the above expression is a ferromagnetic coupling for the core spins
and the second term is an antiferromagnetic coupling. Note that 
$\langle C ^ + _i C _j \rangle _0 $ is only linear on $x$ at small values of the 
electron concentration.

\section{Monte Carlo results with  the approximated 
double exchange energy}

In figure \ref{de-cde} we plot, for $x=0.1$, $x=0.25$ and $x=0.3$, 
the absolute value of the magnetization, $M$, 
as a function of the temperature, $T$,  for the case where the internal energy is
obtained by using the approximated DE energy, equation (\ref{energy-approx}),
for comparing with the case in which it is obtained by diagonalizing the
DE Hamiltonian.  
The difference between both curves is very small,
and the corresponding critical temperatures are practically the
same, see figure \ref{tc}. Therefore we conclude that
equation (\ref{energy-approx}) is a good approximation for the energy of the DE model.

Using equation (\ref{energy-approx}),  for the DE energy, we can perform MC simulations 
in much bigger systems, and obtain more precise values of the critical temperatures.
In figure \ref{N=20} we plot the magnetization 
as a function of the temperature for a sample of size $N=20$.
For this size the critical temperature is obtained with a
big precision. We have checked that for $N=20$ the values  
of the critical temperatures are accurate in two digits and
that it is not worthwhile to increase more the size
of the system.
We have performed MC simulations for different values of the electron concentration
and, in figure \ref{tc},
we have plotted the magnetic phase diagram.

Kubo and Ohata\cite{kubo} have developed  an expression for the
DE energy which coincides with $E_0$. 
In order to know the importance of  $E_2$ in the calculation
of the value of the
critical temperature, 
we have also computed the value of $M$ as a function of $T$ 
by using only $E_0$ for the internal energy, see  figure \ref{N=20}.
Comparing these results with the ones obtained
with the approximated DE energy, $E_0+ E_2$,
we estimate that the inclusion of $E_2$ in the
internal energy lowers the critical temperature more than a 20$\%$.

In Fig.\ref{tprome}, we plot for different values of the concentration $x$, 
the statistical average
of the absolute value of the hopping amplitude, 
$ \left | t_{i,j} \right |   $, as a function of the 
temperature. This quantity is
proportional to the electron bandwidth. We find that 
$\langle \left | t_{i,j} \right | \rangle$ is 
a continuous function of $T$, and there are not signals of any change in the 
electronic structure near $T_c$.

The bandwidth near $T_c$ is  around 1.15 times bigger than the bandwidth 
in the $T \rightarrow \infty$ limit. Li {\it et al}\cite{li}
have obtained that in the $T \rightarrow \infty$ limit of the 
cDE model, only $\sim 0.5 \%$ of the electron states are localized.
In order 
to know the difference between the density of states at 
$T \sim T_c$ and at $T \rightarrow \infty$, we have calculated the 
statistical average of the electron density of states. 
We find that the number of electron states with energies between 
$-W(T_c)$ and 
$-W(T \rightarrow \infty )$ is less than $1 \%$ of the total number of states. 
Therefore, even if all these states were localized, the percentage 
of localized states near $T_c$ is not larger than $2 \%$. We conclude
that in the DE models it is not possible to relate
the metal-insulator transition with the ferro-paramagnetic
one.

The critical temperatures we obtain are around 1.5 times smaller than 
the obtained in mean field theory by Kubo and Ohata\cite{kubo}.
However, it is around 8 times smaller than the obtained 
by Millis {\it et al.}\cite{millis1} using also a mean field approximation.
We think that the  difference occurs because they use a lineal dependence,
see Fig. \ref{coef},  of the 
kinetic electron energy which overestimate the value of $\langle C_i ^+ C_j\rangle_0$,
and therefore the value of $T_c$.

From Kubo and Ohata model (\ref{tmf}), it is clear that $T_c$ scales
with $t\langle C_i ^+ C_j\rangle_0$ and the magnetization versus $T/T_c$
curve is independent of the electron concentration x. 
But there is no reason to think
that this must be the case when the second term, $E_2$, of our approach is
included in the calculations. However, we have found that 
the $M$ versus $T/T_c$ curves
almost coincide (see Fig.\ref{N=20}). 
This fact is not only found in the frame of our approximation
to the DE energy but also when the DE and cDE Hamiltonians
are diagonalized.

The presented results are in qualitatively agreement with those obtained by Dagotto
{\it et al}\cite{dagotto}. However our estimations of $T_c$ are $\sim 1.5$  
times bigger.  
The discrepancy is due to the difference in the criterion  used to
obtain $T_c$. In reference \cite{dagotto} the $T_c$ is defined
as the temperature where the spin-spin correlation
function in real space becomes zero at the maximum distance
available in the unit cell. They used a unit cell of size 
$6 \times 6 \times 6 $.
Our criterium is based in the change of sign of the second derivative of
$M$ with respect to $T$. We have checked that both criteria give the same $T_c$
in big unit cells, but the criterium based on the correlation
function underestimate $T_c$ in small unit cell calculations. 
This is clear in Fig.\ref{correlation} where we plot the 
spin-spin correlation function versus
distance for several temperatures and  for two unit cells with
$N=6$ and $N=10$. The electron concentration is $x=0.14$ to compare with the 
results presented in reference \cite{dagotto}.
In the case of $N=6$ it seems that the critical temperature 
is smaller than $T=1/15$, because the correlation function is zero 
at a separation of 8$a$. However, when $N=10$ it is clear that 
the ferromagnetic correlation at $T=1/15$ is not zero at 8$a$,
and in fact it seems that at this temperature the system is still
ferromagnetic. 
On the other hand, our criterium overestimate the $T_c$ for small
unit cells and the possibility of increasing the size leads to more realistic
values of the critical temperature. 

To compare with the experimental critical temperatures, it is necessary an estimation
of $\bar t$. The experimental bandwidth is  around 1-4eV\cite{rvw}, and 
therefore the value of $\bar t$  is around 0.08-0.3eV. With this result our estimate of the
critical temperature is between 150K and 500K at $x \sim 0.2$. This 
value of $T_c$ is in the range of the observed experimentally. 

\section{Conclusions}
Using Monte Carlo techniques, we have obtained the magnetic phase diagram
of the double exchange Hamiltonian. Comparing the results obtained from 
the double exchange Hamiltonian with a complex hopping (cDE) and 
with its absolute value (DE) we have found that the Berry phase 
of the hopping has a negligible effect in the magnetic critical temperature.
This implies that it is possible to choose wave functions phase factors
that counteract the phase of the hopping such that it is not relevant 
in the MC calculations. 
To avoid the limitations on the size of the 
systems studied we have developed a second order perturbative approach 
to calculate the electron kinetic energy without diagonalizing
the Hamiltonian.
Within this approach we have calculated the 
critical temperature for a bigger size of the system ($N=20$)
and, consequently, more accurately.
The values of $T_c$ obtained are in the range of the 
experimental ones and in qualitative agreement with those in 
ref. \cite{dagotto}.
Neither the average of the electron density of states
nor the kinetic energy
reveal any strange behavior near the 
magnetic critical temperature. We conclude that,
in order to understand the metal-insulator transition
which occurs in the manganites,
another effects, not included in the double
exchange  Hamiltonian, should be taken into account.

{\it Acknowledgments.}
This work was supported by the CICyT of Spain under Contract No. PB96-0085,
and by the Fundaci\'on Ram\'on Areces.
Helpful conversations with F. Guinea, R. Ram\'{\i}rez, S. Das Sarma and J.A. Verg\'es
are gratefully acknowledged.

\begin{figure}
\caption{
Critical temperature of the ferro-paramagnetic transition versus concentration
of conduction electrons for different approaches. Squares correspond to the 
Hamiltonian (\ref{hcomplex}) with the hopping given by (\ref{tcomplex}); circles
to hopping (\ref{treal}). The curves are related to the calculations made with 
an approximated expression obtained in order to avoid diagonalization in the 
MC simulations (ADEE: approximation to the double exchange energy).
This fact allows us to increase the size of the system.
The line is only a guide to the eye. The critical temperatures obtained
are in the range of the experimental ones. 
}  

\label{tc}
\end{figure}

\begin{figure}
\caption{
Absolute value of the magnetization versus the temperature for 
concentrations $x=0.1$, $0.25$ and $0.3$. The size of the
unit cell is $N=4$. 
In order to compare we have plotted also the curves obtained
with the cDE and DE Hamiltonians as well as with the perturbative
approach developed here (ADEE). 
}
\label{de-cde}
\end{figure}

\begin{figure}
\caption{
Double exchange energy versus temperature. The lack of difference in these curves
implies the result in Fig. \ref{de-cde}. 
The corresponding critical temperatures are pointed with an arrow. 
}
\label{E-t}
\end{figure}

\begin{figure}
\caption{
Bandwidth (a) and chemical potential (b) versus temperature. 
In the disordered case, $T \rightarrow \infty$,
$W$ should be equal to $8$. Here, in the paramagnetic phase, $W>8$ because the 
transition is second order (see text for details). Note that 
the curves are continuous near $T_c$ which is pointed with an arrow
in each case.
} 

\label{W-mu-t}
\end{figure}

\begin{figure}
\caption{
Coefficients used in the perturbative approach to the DE Hamiltonian versus
concentration (see eq. (\ref{energy-approx})).
}
\label{coef}
\end{figure}

\begin{figure}
\caption{
Magnetization as a function of $T/T_c$ as obtained using 
the $0^{th}$ order approximation (dashed line), $E_0$ in eq. (\ref{energy-approx}),
and the $2^{nd}$ order approximation (continuous line) to the double exchange energy
(ADEE),
eq. (\ref{energy-approx}). The critical temperature taken
for reference is the one that corresponds to the ADEE model. 
In the case of the $0^{th}$ order approximation, the $M$ versus $T/T_c$
curve is independent of the electron concentration. In the ADEE model,
there is no reason for this independence; however, we have found 
that $M(T/T_c)$ is practically independent of x. To show this,
the $M(T/T_c)$ curve has been plotted for various concentrations
($x=0.1$, $0.2$, $0.3$, $0.4$ and $0.5$).
}
\label{N=20}
\end{figure}

\begin{figure}
\caption{
Average of the absolute value of the hopping amplitude,
$\langle \left | t_{i,j} \right | \rangle$,  versus temperature for $N=20$ and
concentrations $x=0.1$, $0.25$ and $0.3$. This average
is proportional to the bandwidth. 
We recover the continuous behavior that we found on Fig.4.  
}
\label{tprome}
\end{figure}

\begin{figure}
\caption{
The spin-spin correlation versus distance is plotted for different
temperatures. In (a) $N=6$ and in (b) $N=10$. For these calculations
we have used the perturbative approach. If the criterium to obtain
$T_c$ is that the spin-spin correlation becomes zero, a smaller
size of the system leads to an underestimation of the critical temperature. 
}
\label{correlation}
\end{figure}

\end{document}